\pdfoutput=1
\documentclass[aip,reprint,floatfix]{revtex4-2}

\usepackage{amsmath}
\usepackage{graphicx}
\usepackage{hyperref}

\begin{document}


\title{Mechanisms for the Near-UV Photodissociation of CH$_3$I on D$_2$O/Cu(110)} 



\author{E.R. Miller}
\author{G.D. Muirhead}
\author{E.T. Jensen}
\email[email:]{ejensen@unbc.ca}
\affiliation{Physics Department \\ University of Northern British Columbia \\ 3333 University Way, Prince George B.C. V2N 4Z9, Canada}


\date{August 22, 2012}

\begin{abstract}
The system of CH$_3$I adsorbed on submonolayer, monolayer and multilayer thin films of D$_2$O on Cu(110) has been studied by measuring the time-of-flight (TOF) distributions of the desorbing CH$_3$ fragments after photodissociation using linearly polarized $\lambda$=248nm light. For multilayer D$_2$O films (2--120ML), the photodissociation is dominated by neutral photodissociation via the ``A-band'' absorption of CH$_3$I. The polarization and angle dependent variation in the observed TOF spectra of the CH$_3$ photofragments find that dissociation is largely via the $^3Q_0$ excited state, but that also a contribution via the $^1Q_1$ excitation can be identified. The photodissociation results also indicate that the CH$_3$I adsorbed on D$_2$O forms close-packed islands at submonolayer coverages, with a mixture of C--I bond axis orientations. For monolayer and submonolayer quantities of D$_2$O we have observed a contribution to CH$_3$I photodissociation via dissociative electron attachment (DEA) by photoelectrons. The observed DEA is consistent with delocalized photoelectrons from the substrate causing the observed dissociation-- we do not find evidence for an enhanced DEA mechanism via the temporary solvation of photoelectrons in localized states of the D$_2$O ice.
\end{abstract}


\maketitle 

\section{Introduction}

Interest in electron and UV stimulated chemical processes on and inside icy materials comes from a wide variety of scientific disciplines. In planetary science and astrophysics, these systems have been invoked as possible routes to the formation of molecules necessary for the building blocks of life \cite{Bernstein:2002tc,MunozCaro:2002iw,MunozCaro:2003gt}. These systems are also proposed as a mechanism for the generation of ozone depleting species via processes that occur on stratospheric ice particles\cite{Lu:2010fu}. Electron solvation in water\cite{Jordan:2004ea} and on water clusters\cite{Jordan:2010up} has been an ongoing theme in physical chemistry with a very large existing literature. Enhanced stimulated desorption and dissociation of a variety of species coadsorbed with water ice has been studied in a variety of ways for several years now\cite{Lu:1999ta,Lu:2004eb,Perry:2007db,Ryu:2006cy,Sohn:2009tv}. In this context, methyl iodide is an interesting system to study since it is created in substantial quantities in the Earth's oceans and is the most abundant terrestial organoiodide, playing a role in ozone chemistry and the formation of atmospheric aerosols\cite{Smyth:2006,Solomon:1994}. Gas-phase methyl iodide possesses large cross sections for dissociation by near-UV light as well as by low energy electrons. The dissociation and reaction of CH$_3$I on D$_2$O thin films has been studied in some detail using an external electron beam (energies between 180 and 210 eV)\cite{Perry:2007db}. In that study, the role of the primary electron beam and that of low energy secondary electrons as principal mechanisms was discussed in detail for the observed dissociation, desorption and chemical reaction pathways. The overall cross section for CH$_3$I decomposition was found to be comparable to gas phase values for the electron energies used. A more recent study\cite{Sohn:2009tv} utilizing TPD, UPS and XPS as a probe of CH$_3$I/D$_2$O/Cu(111) photochemistry using near-UV photon irradiation from an Hg lamp source (having large photon intensities near 254nm and also at 290--330nm) reported evidence for a significantly enhanced dissociation of CH$_3$I. The enhanced dissociation was ascribed to a facile electron transfer from electrons solvated at the ice surface, the ice being an important intermediate by efficiently capturing photoelectrons emitted from the underlying metal substrate.

Details of the temporary capture of photoelectrons by thin ice films on metals has recently emerged from femtosecond two-photon photoemission (fs-2ppe) experiments. These works have shown that for a variety of thin water films on metal surfaces, a short-lived surface solvated electron state exists. In the case of D$_2$O/Cu(111) the solvated electron state is found at an energy of $\sim$2.9eV above the Fermi level\cite{Staehler:2008cx}, although the energies shift as a function of substrate, ice structure and time-- these states can decay over timescales ranging from a few hundred femtoseconds to minutes\cite{Bovensiepen:2009kj}. 

These surface solvated electrons are attractive as a mechanism for dissociation of a variety of halocarbon species, as many members of this class of molecules possess substantial dissociative electron attachment (DEA) cross sections for low energy electrons. That transient surface bound photoelectrons can cause DEA of coadsorbed halomethanes was recently demonstrated in our laboratory\cite{Jensen:2008jb}, though in this case the electrons were in delocalized image states above {\it n}-hexane, while the solvated surface electrons on D$_2$O ices are in localized states. The transfer of electrons solvated in D$_2$O thin films to coadsorbed chlorocarbons has been observed using fs-2ppe techniques\cite{Ryu:2006cy,Bertin:2008kw} as well as studied theoretically\cite{Bhattacharya:2010cb,Fabrikant:2012hw}.

Methyl iodide is known to have a very large DEA cross section for low energy electrons-- both from gas phase studies\cite{Wilde:2000ul} and also from low energy electron beam studies for CH$_3$I in the adsorbed state\cite{Jensen:2008jv}. Although the vibrational Feshbach resonance responsible for part of the extremely large gas-phase DEA cross section at low energies is quenched\cite{Jensen:2008jv,Fabrikant:2011ep} for adsorbed CH$_3$I, the DEA cross section for CH$_3$I adsorbed on 10ML of Kr ($\sigma=6.8\times 10^{-16}cm^2$)\cite{Jensen:2008jv} at low energies exceeds that for adsorbed CH$_3$Br ($\sigma=3.0\times 10^{-16}cm^2$)\cite{Ayotte:1997th}. However, external electron beams are restricted to incident energies above the vacuum level, while cross sections in the adsorbed state for these molecules are likely largest for electron energies between the Fermi and vacuum levels, due to the downward shift of the affinity level caused by the molecular anion's polarization interactions with the surface region. This range of electron energies can be accessed by generating photoelectrons from the metal using near UV light. In this way, a broad energetic range of photoelectrons is created\cite{Weik:1993wa}, and those electrons that migrate to the surface region and have the requisite energy can then interact with the adsorbed species. The ability to modify the nascent photoelectron distribution by adsorption of various molecular thin films is an attractive prospect, particularly in studies of coadsorbates possessing electron affinity levels in this energy region.

Gas-phase methyl iodide also has a large cross section for photodissociation in the near-UV photon energies-- the ``A-band" absorption extends roughly from 220nm to 300nm, with a peak cross section of $1.2\times 10^{-18}cm^2$ at 258nm\cite{KellerRudek:2013wf}. Neutral photodissociation of CH$_3$I has been studied extensively in the gas-phase\cite{Eppink:1998ue} as well as a variety of surface contexts\cite{Fairbrother:1994vj,Srivastava:2002wi,Jensen:2005js}, including recent work using fast time-resolved techniques for CH$_3$I on MgO and gold surfaces\cite{Vaida:2008gv,Vaida:2009kh,Vaida:2011gb}. Absorption of photons in the A-band can lead to prompt scission of the C--I bond. Due to the rather large spin-orbit coupling in Iodine, there is a substantial difference in energy ($\Delta$E=0.943eV\cite{Eppink:1998ue}) between ground state iodine (CH$_3$I$\rightarrow$CH$_3$ + I($^2P_{3/2}$)) and electronically excited iodine (CH$_3$I$\rightarrow$CH$_3$ + I*($^2P_{1/2}$)), leading to differences in the translational energy partitioned to the CH$_3$ moeity in the two different pathways. Detailed studies of gas-phase photodissociation found that the A-band is composed of overlapping Franck-Condon transitions to three excited states, labeled (in Mulliken's notation) $^3Q_1$, $^3Q_0$ and $^1Q_1$ in order of increasing energy. The transition $X-{^3Q_0}$ dominates the A-band absorption and the transition moment for this excitation is parallel to the C--I bond axis. The transitions $X-{^3Q_1}$ (lower energy) and $X-{^1Q_1}$ (higher energy) are weaker and have transition moments perpendicular to the C--I bond axis. A complication in the analysis of A-band CH$_3$I photodissociation is the presence of a curve crossing just beyond the Frank-Condon region, between the $^3Q_0$ and $^1Q_1$ states. A substantial fraction of the molecules photoexcited to the $^3Q_0$ state can `hop' to the $^1Q_1$ surface due to the effects of spin-orbit coupling of I and symmetry breaking for the CH$_3$I on the $^3Q_0$ potential energy surface\cite{Amatatsu:1991}. The result is that even for molecules excited exclusively to the $^3Q_0$ state, which asymptotically corresponds to the I*($^2P_{1/2}$) pathway, dissociation can also proceed via the I($^2P_{3/2}$) pathway. Experiments in the gas-phase or from adsorbed CH$_3$I find varying proportions of dissociating molecules following the I and I* paths. In favourable circumstances, photodissociation experiments on CH$_3$I that is adsorbed on surfaces with a fixed orientation can utilize the vector nature of the A-band absorptions in order to untangle the dynamics\cite{Jensen:2005js}.

\section{Experimental Details}
The experiments were performed in an ultra-high vacuum system that has been described previously\cite{Jensen:2005js}. The Cu(110) single crystal sample is cooled by liquid nitrogen (base temperature 90K) and can be heated by electron bombardment to 920K for cleaning. Sample cleanliness and order were monitored by Auger electron spectroscopy and low energy electron diffraction measurements respectively. Neutral products from surface photodissociation travel 185mm to pass through a 4mm diameter aperture to a differentially pumped Extrel quadrupole mass spectrometer with an axial electron bombardment ionizer. The sample to ionizer distance is 203mm. Ions created in the ionizer then travel through the quadrupole region and are mass selected, in the present experiments using m/q=15amu. Ion arrivals are recorded using a multichannel scaler that begins counting 50$\mu s$ prior to the initiating laser pulse, and the counts from multiple laser pulses are summed. All of the spectra shown in the present work are the result of summing data from 1000 laser pulses into 1000 1$\mu s$ time bins.

The laser pulses ($\sim$5ns duration) are produced by a small excimer laser (MPB PSX-100) operating at 20Hz. Linearly polarized light has been used exclusively in this work. To create polarized light, the beam passes through a birefringent MgF$_2$ crystal to separate p- and s-polarized components, which can then be directed at the sample. Unless otherwise stated, p-polarized laser pulses were used in the present study. The s-polarized light was derived from the p-polarized beam by inserting an antireflection coated zero order half waveplate into the beam, which preserved the laser alignment on the sample with minimal absorption and reflection losses. In this work 248nm (KrF) laser light was used, with laser fluences on the sample of $\sim$ 0.8mJ/cm$^2$ per pulse. The laser pulses were collimated using a 6mm diameter aperture and were unfocussed on the sample. The laser light is incident upon the sample at a fixed angle of 45$^\circ$ from the mass spectrometer axis-- for example, when the Cu(110) sample is oriented to collect desorption fragments along the surface normal direction, the light is incident at 45$^\circ$.

For these experiments a microcapillary array directed doser has been added to the UHV system, following the design of \textcite{Fisher:2005uw}. Deposition of molecular films is done with the sample held normal to the doser, 25mm away. This was found to enhance the deposition by a factor of 10 compared to background dosing. In the present work, all dosing was done with the sample close to the base temperature of 90K. All gas doses are reported as uncorrected ionization gauge readings. The CH$_3$I used in this work (Aldrich 99.5\%) was transferred via a glass and teflon gas-handling system and degassed by freeze/pump/thaw cycles. The D$_2$O (Aldrich, 99.9 atom \% D) used in this work was degassed by multiple freeze/pump/thaw cycles and was contained in a pyrex vial a few cm from the precision leak valve used to admit gases to the directed doser. The CH$_3$I dosing was calibrated by observation of the photochemical behaviour of CH$_3$I/Cu(110)-I, in which a pronounced change in photochemical activity is observed at the transition from the first to second monolayer of CH$_3$I. In this case, 0.93$\pm$0.02L CH$_3$I was found to correspond to 1.0ML for the Cu(110)-I substrate. In the present case of adsorption of CH$_3$I on D$_2$O thin films in which the surface structure is less well characterized and no distinct TPD or photochemical signatures define what dose corresponds to monolayer CH$_3$I, we report CH$_3$I doses in terms of effective monolayers based on this calibration (1.0ML=0.93L). For D$_2$O we determined an effective monolayer calibration based on findings from TPD of D$_2$O/Cu(110) and titration of CCl$_4$ on top of varying D$_2$O films, which formed atomic chlorine on the metal surface after warming to desorb the molecular layers. From this we found 1.0ML=0.30L for D$_2$O, with somewhat larger errors bars than was the case for CH$_3$I. Adsorption of D$_2$O on Cu(110)-I is less well characterized, as the TPD features do not reveal clear information on D$_2$O layer completion, and titration of Cl from CCl$_4$ is not possible. For the present, we have used the same dose calibration as for the clean Cu(110) surface. In the case of D$_2$O multilayers, we expect only small differences in overall coverage between the clean and iodized Cu(110) substrates.

\section{Observations and Discussion}

In the thin film experiments reported here (D$_2$O doses up to 120ML), the D$_2$O was adsorbed on the Cu(110) substrate at 90K. Under the deposition conditions of our system, we believe that all of the multilayer D$_2$O thin films were low porosity amorphous solid water (ASW)\cite{Stevenson:1999ut}\footnote{We attempted to create higher porosity D$_2$O films using high angle of incidence deposition\cite{Stevenson:1999ut} but all of the photodissociation experiments using various D$_2$O multilayer films (varying the angle and temperature of D$_2$O deposition) exhibited essentially the same behaviour for photodissociation of CH$_3$I. Based on our observations and the data of Fig. 3 in \textcite{Stevenson:1999ut}, we believe that our apparatus was not able to grow D$_2$O films with significant porosity. While we do not believe that the D$_2$O films studied are of uniform thickness on the Cu(110) surface\cite{Yamada:2006hb}, we do believe that for the thicker films studied, that the Cu(110) surface is effectively covered by D$_2$O.}. Adsorption of CH$_3$I on top of D$_2$O/Cu(110) was studied using temperature programmed desorption, which identified two CH$_3$I desorption features. Our TPD findings are in accord with those of Ref. \onlinecite{Perry:2007db} who studied CH$_3$I on D$_2$O/Ru(0001). At low doses an adsorbed CH$_3$I species that desorbs at $\sim$125K for $\lesssim$0.33ML CH$_3$I (the $\alpha$ state) is present. As the CH$_3$I dose is increased, a higher desorption temperature ($\sim$135K) feature begins to grow (the $\beta$ state), and this gradually becomes the dominant feature and is the only TPD feature present above 1.0ML CH$_3$I dose. 

\begin{figure}
\includegraphics[scale=0.60]{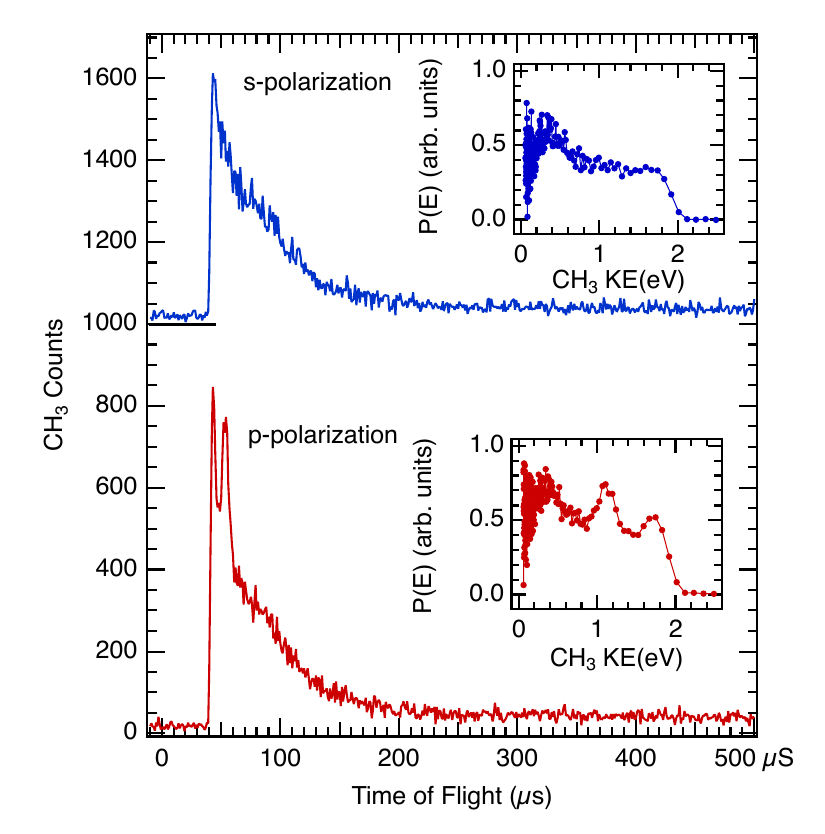}
\caption{\label{Fig_p_s_pol_tof} Time of flight spectra from the photodissociation of submonolayer (0.33ML) CH$_3$I on 15ML D$_2$O/Cu(110) using $\lambda$=248nm light. The CH$_3$ photofragments are collected in the surface normal direction, while the light is incident at 45$^\circ$ from normal. The spectrum from p-polarized light (lower trace) shows two ``fast'' peaks from dissociation via the $^3Q_0$ excited state of CH$_3$I, with a tail of lower energy CH$_3$ fragments to $\sim 250\mu$s flight time. The inset plots show the data plotted as the flux versus translational energy. }
\end{figure}

 The photon energy used in the present work ($h\nu$=5.0eV) is less than the bandgap energy of 7.5eV for water ice\cite{Kobayashi:1983ww} so the thin D$_2$O layers used are essentially transparent to the laser light. This photon energy is larger than the work function of clean Cu(110) ($\Phi$=4.48eV)\cite{CRC:2011} so we anticipate the formation of both free and bound (hot) photoelectrons in the surface region. Time-of-flight data from 0.33ML CH$_3$I on multilayer (15ML) D$_2$O thin films is shown in Fig. {\ref{Fig_p_s_pol_tof}}. These spectra are obtained with the QMS detector in the surface normal direction ($\theta$=0$^\circ$) with the light incident at 45$^\circ$ to the normal. Using p-polarized light (lower trace), two fast peaks (flight times of 43$\mu s$ and 53$\mu s$ respectively) are evident, plus a substantial ``tail'' of slower CH$_3$ photodissociation products extending to $\sim$250$\mu s$ flight time. The two fast photodissociation peaks correspond to neutral photodissociation via the I($^2P_{3/2}$) and I*($^2P_{1/2}$) pathways respectively. The spectrum obtained using s-polarized light (upper trace in Fig. {\ref{Fig_p_s_pol_tof}}) exhibits a single fast onset peak at 43$\mu s$ plus a substantial tail of slower CH$_3$, extending to $\sim$250$\mu s$ flight times. Inset plots of the same data are also shown in Fig. {\ref{Fig_p_s_pol_tof}}, in which the data has been converted to show CH$_3$ flux vs. translational energy. The peaks at 1.1eV and 1.75eV correspond to the fast neutral photodissociation features, while the slower broad feature near 0.4eV translational energy corresponds to the shoulder seen in the TOF data. Both the p- and s-polarization TOF spectra can be understood on the basis of neutral photodissociation of the CH$_3$I and the experimental geometry used for the data of Fig. {\ref{Fig_p_s_pol_tof}}. As discussed in the introduction, at $\lambda$=248nm CH$_3$I can dissociate via the $^3Q_0$ excited state with large cross section, however this requires a component of the incident light polarization to be coincident with the CH$_3$I molecular axis. Prompt scission of the C-I bond in this case will lead to dissociation via both the I and I* pathways. Molecules oriented with the CH$_3$ moiety pointing normal to the surface can efficiently absorb p-polarized light to reach the $^3Q_0$ excited state and follow this pathway, with the CH$_3$ photofragments directed toward the QMS detector. Incident s-polarized light can also be absorbed to reach the $^3Q_0$ state, but these molecules will not be pointing the C--I bond axis in the surface normal direction, rather these CH$_3$ fragments will be directed closer to the surface plane. Scattering of these CH$_3$ photofragments can result in particles travelling toward the QMS detector, though with a degraded translational energy distribution. A secondary pathway for photodissociation via excitation to the $^1Q_1$ excited state (utilizing light polarized perpendicular to the molecular axis) can also lead to fast dissociation for CH$_3$I pointing the CH$_3$ moiety in the surface normal direction for s-polarized light, but the cross section for this excitation is likely $\sim$8x smaller than the $^3Q_0$ cross section at this wavelength\cite{Jensen:2005js}. A portion of the fast onset peak at 43$\mu s$ flight time in the s-polarized TOF spectrum of Fig. {\ref{Fig_p_s_pol_tof}} (upper trace) is due to excitation via the $^1Q_1$ pathway. In the present case, it appears that the incident s-polarized light leads to significant contributions from both the $^1Q_1$ and the $^3Q_0$ excitations for molecules at various orientations with respect to the surface normal. Likewise for the p-polarization, the TOF spectrum in Fig. {\ref{Fig_p_s_pol_tof}} (lower trace) displays a large inelastic `tail' that indicates many of the dissociating molecules produce CH$_3$ photofragments that suffer inelastic collisions before being detected. For either polarization of light, the majority of the CH$_3$ fragments detected in the surface normal direction are scattered inelastically, with flight times extending to $\sim$250$\mu s$ and the peaks due to elastic CH$_3$ escaping the surface are superimposed on this substantial background.
 
\begin{figure}
\includegraphics[scale=0.48]{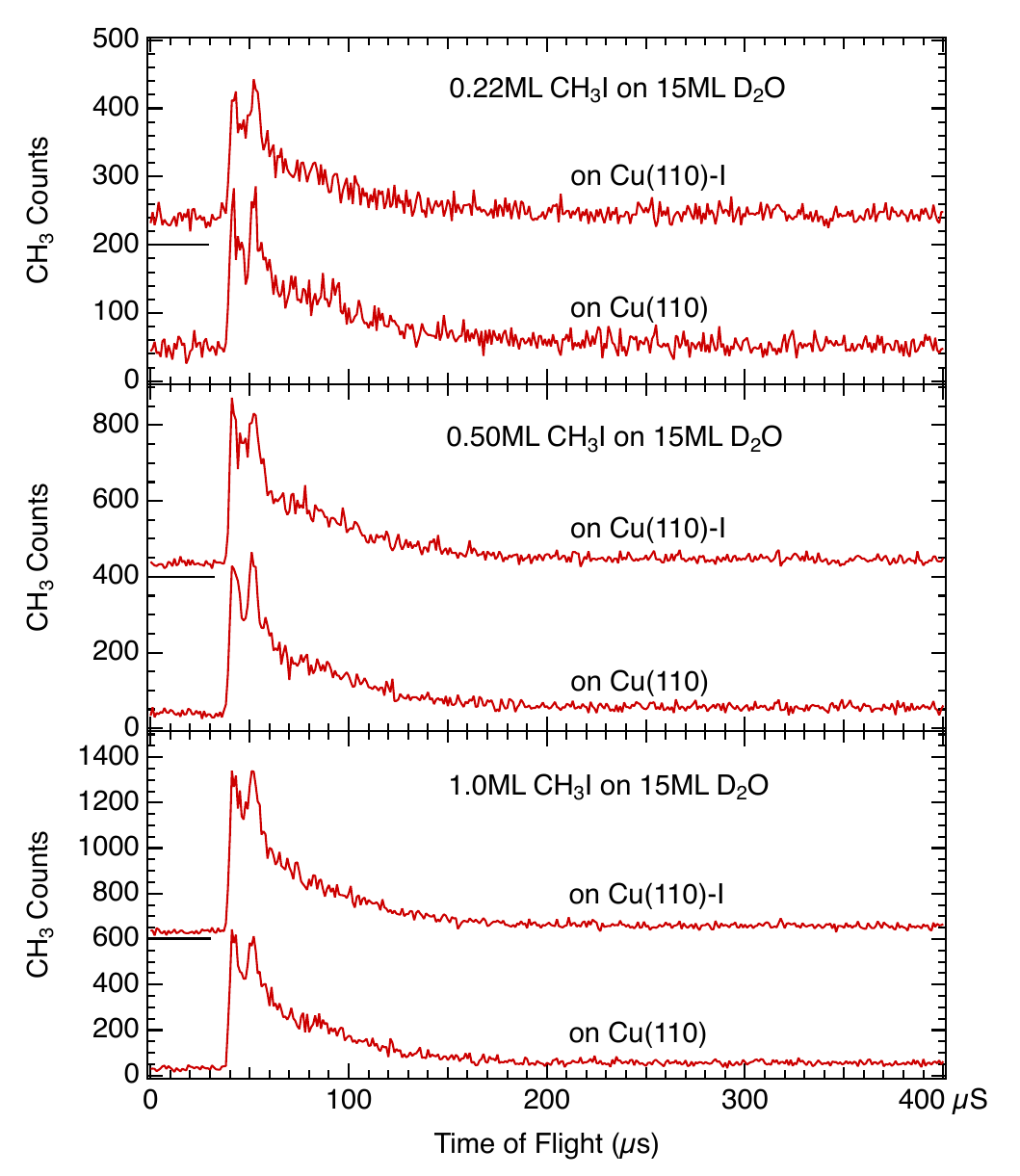}
\caption{\label{Fig_tof_comparisons} Time of flight spectra from varying amounts of CH$_3$I adsorbed on 15ML D$_2$O/Cu(110) and D$_2$O/Cu(110)-I. Each spectrum is obtained from 1000 pulses of p-polarized 248nm laser light, detecting the CH$_3$ photofragments in the surface normal direction. In each case, the signals from the clean and iodized Cu(110) substrates are essentially identical.}
\end{figure}

The photodissociation of varying amounts of CH$_3$I on 15ML D$_2$O thin films is shown in Fig. {\ref{Fig_tof_comparisons}}. The overall yield of CH$_3$ photofragments increases linearly with CH$_3$I dose up to 1.0ML, and the overall TOF spectrum distribution does not change significantly with CH$_3$I dose. The lowest CH$_3$I dose shown in Fig. {\ref{Fig_tof_comparisons}} (0.22ML, upper trace) corresponds to the $\alpha$ state identified in the TPD data. Higher doses of CH$_3$I in which both the $\alpha$ and $\beta$ adsorption states are populated (0.5ML, middle trace), or predominantly the $\beta$ state (1.0ML, bottom trace) produce TOF spectra that have higher intensity but essentially the same profile with no new dynamical channels visible in the data and having the same shape of inelastic ``tail'' in the TOF spectra. It is also significant that the CH$_3$ photodissociation yields in Fig. {\ref{Fig_tof_comparisons}} are essentially the same whether a Cu(110) or the iodized Cu(110)-I substrate is used. The use of the c(2x2) Cu(110)-I substrate is known to suppress photoelectron emission, including the subvacuum level `hot' photoelectrons that are likely most efficient for DEA of the adsorbed CH$_3$I. The same observations have been made using a chlorinated Cu(110)-Cl substrate (data not shown). The observations from the data of Fig. {\ref{Fig_tof_comparisons}} show that for these relatively thick D$_2$O multilayers, dissociation of the CH$_3$I on D$_2$O is not significantly aided by either free photoelectrons or photoelectrons that are solvated at the ice surface and that all of the observed photodissociation is a consequence of neutral photodissociation via A-band pathways. That thick water layers inhibit photoelectron induced dissociation for molecules adsorbed atop is consistent with an earlier study of related systems\cite{Gilton:1989tg}.

\begin{figure}
\includegraphics[scale=0.62]{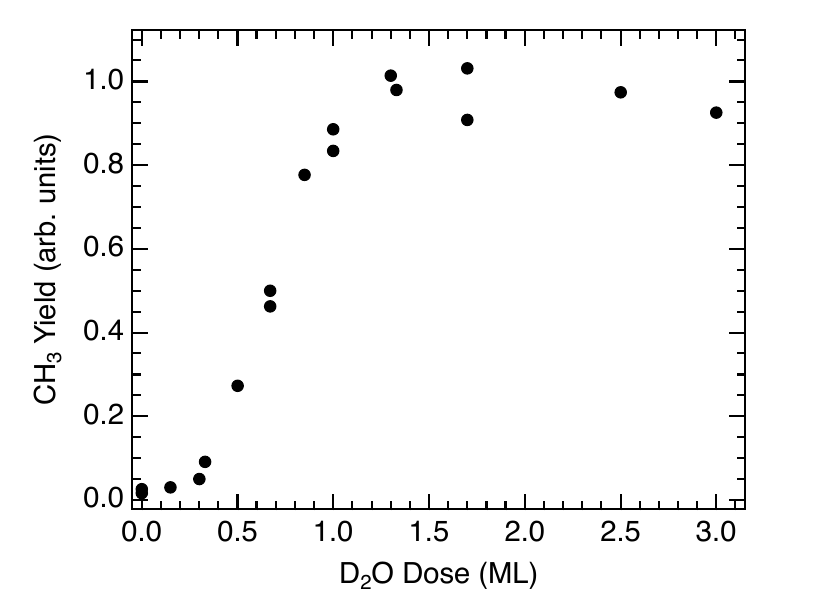}
\caption{\label{Fig_yield_vs_dose} Measured yield of CH$_3$ photofragments from dissociation of 0.33ML CH$_3$I on varying amounts of D$_2$O/Cu(110). These yields were measured in the surface normal direction. For D$_2$O coverages above 3.0ML, the yield of CH$_3$ from photodissociation is essentially constant up to the highest coverages studied (120ML D$_2$O).}
\end{figure}

The yield of CH$_3$ photodissociation products in the surface normal direction for a fixed amount of CH$_3$I on varying doses of D$_2$O/Cu(110) is shown in Fig. {\ref{Fig_yield_vs_dose}}. The photodissociation yields were extracted from TOF spectra between 35$\mu s$ and 230$\mu s$, and correction was made for the differing ionization probabilities of CH$_3$ fragments having differing speeds in the QMS ionizer.

\begin{figure}
\includegraphics[scale=0.65]{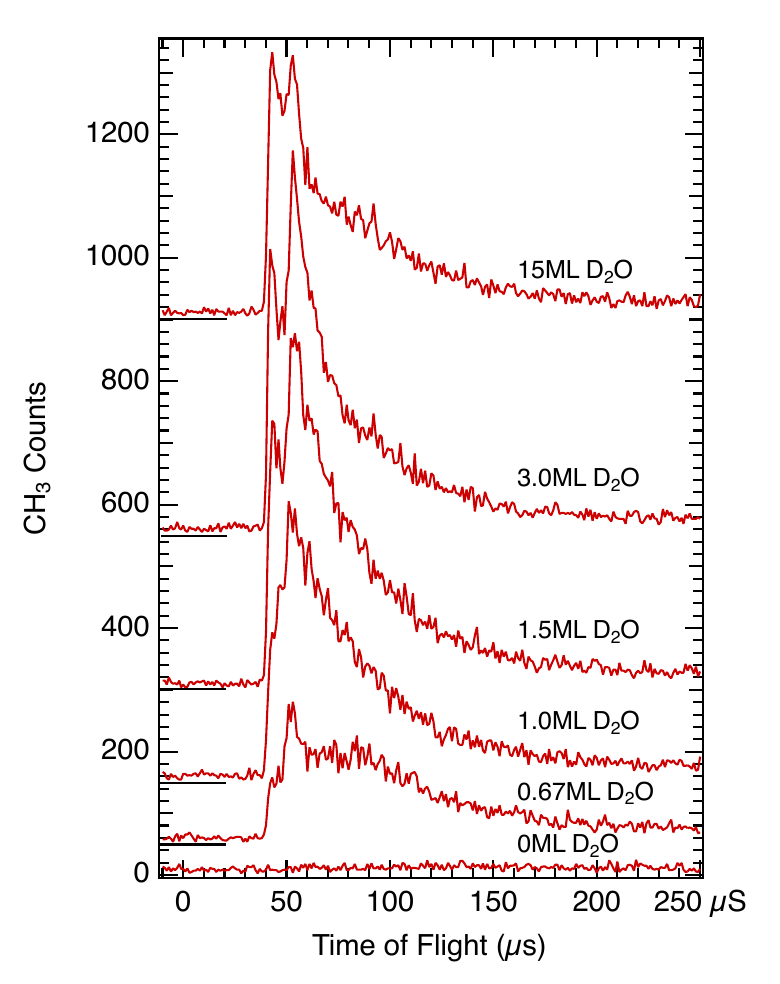}
\caption{\label{Fig_vary_D2O_tof} Time of flight spectra showing photodissociation from 0.33ML CH$_3$I adsorbed on varying thickness films of D$_2$O on Cu(110). The overall yield of CH$_3$ is low until the first layer of D$_2$O is completed on the surface. For thicker D$_2$O films, the CH$_3$ yield and TOF distribution remains essentially constant.}
\end{figure}

\begin{figure}
\includegraphics[scale=0.55]{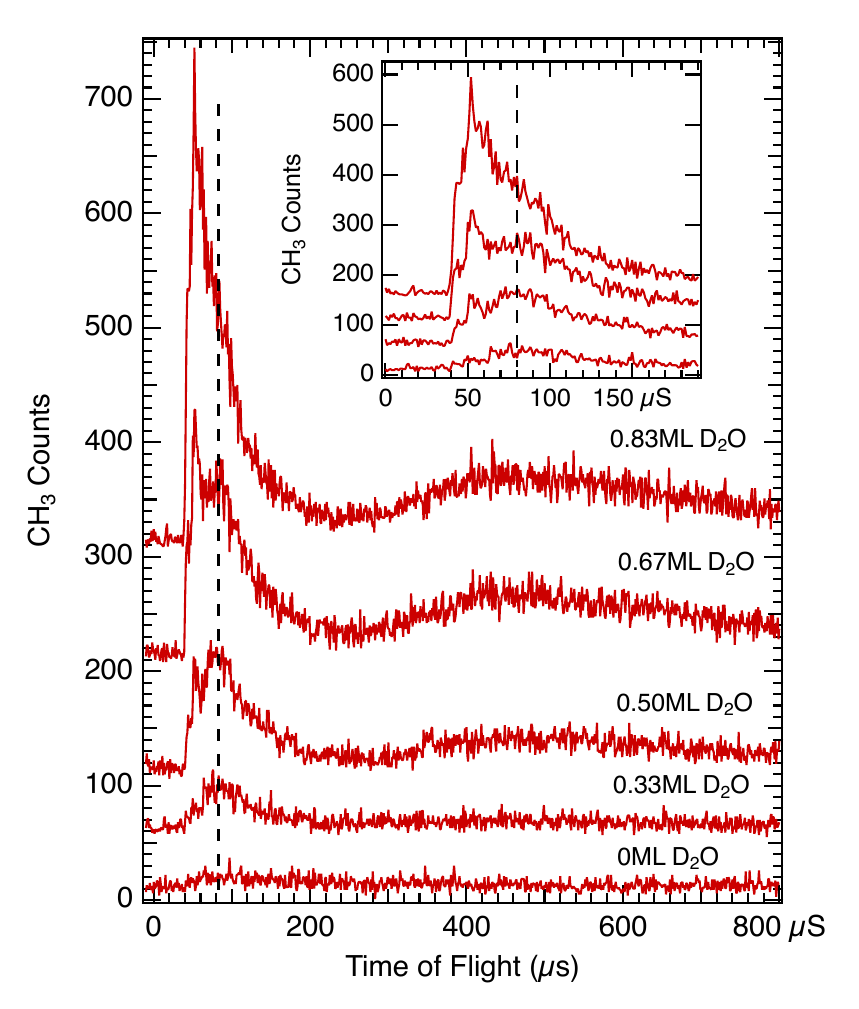}
\caption{\label{Fig_low_dose_tof} Detail of the variation in time of flight spectra for 0.33ML CH$_3$I coadsorbed on a range of submonolayer quantities of D$_2$O on a Cu(110) substrate. A feature due to dissociative electron attachment by photoelectrons is observed near 80$\mu s$ flight time (dashed line), and a feature due to photodesorption of intact CH$_3$I molecules is visible between 250$\mu s$ and 850$\mu s$ flight times. The inset shows the same data for 0.33ML D$_2$O and above, highlighting shorter flight times. }
\end{figure}

\begin{figure}
\includegraphics[scale=0.58]{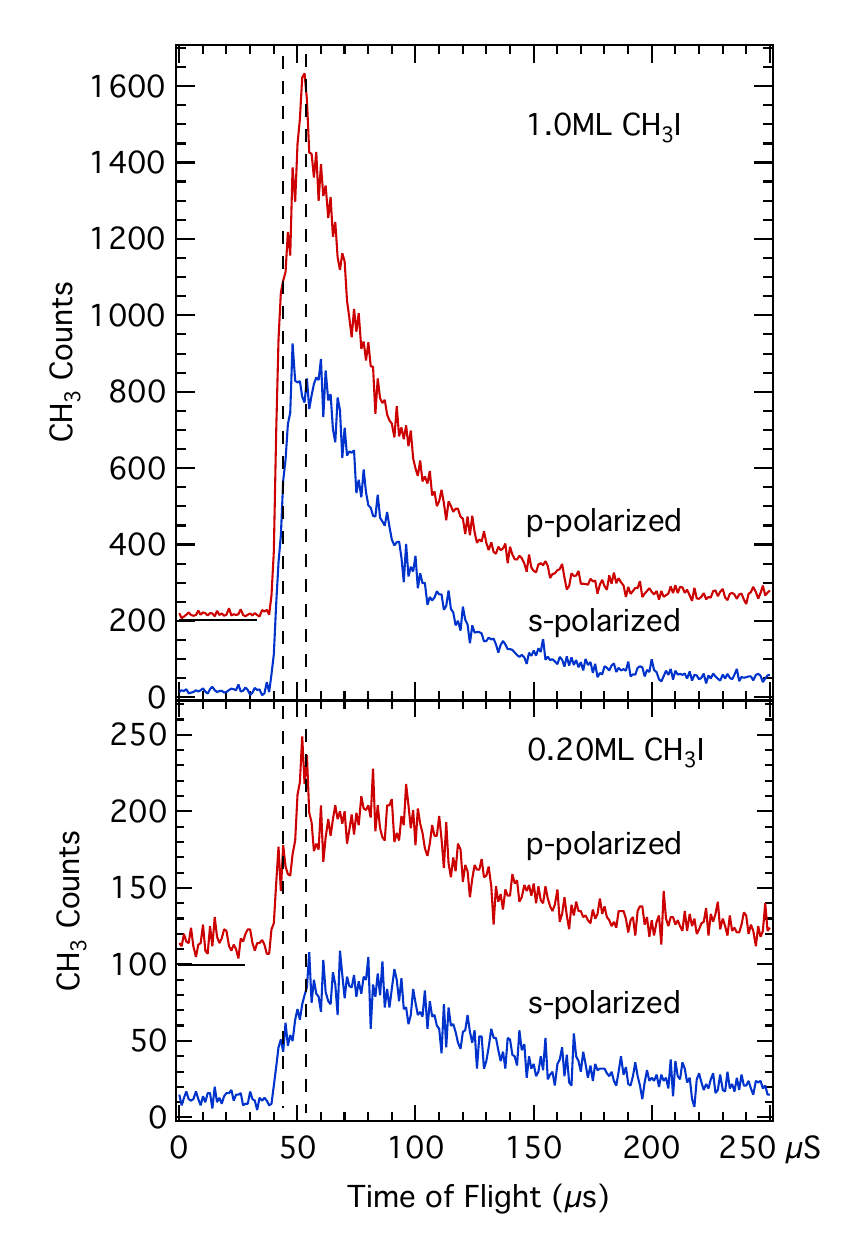}
\caption{\label{Fig_submonolayer_p_vs_s} Comparisons of time of flight spectra for CH$_3$I adsorbed on submonolayer (0.67ML) D$_2$O/Cu(110) obtained in the surface normal direction using p-polarized light and s-polarized light. Photodissociation features for CH$_3$I via the $^3Q_0$ dissociative state (peaks at 43$\mu s$ and 53$\mu s$) are suppressed in the s-polarized data, while the indirect dissociation via photoelectrons is comparable between the two light polarizations (aside from differences due to substrate absorption). }
\end{figure}

\begin{figure}
\includegraphics[scale=0.50]{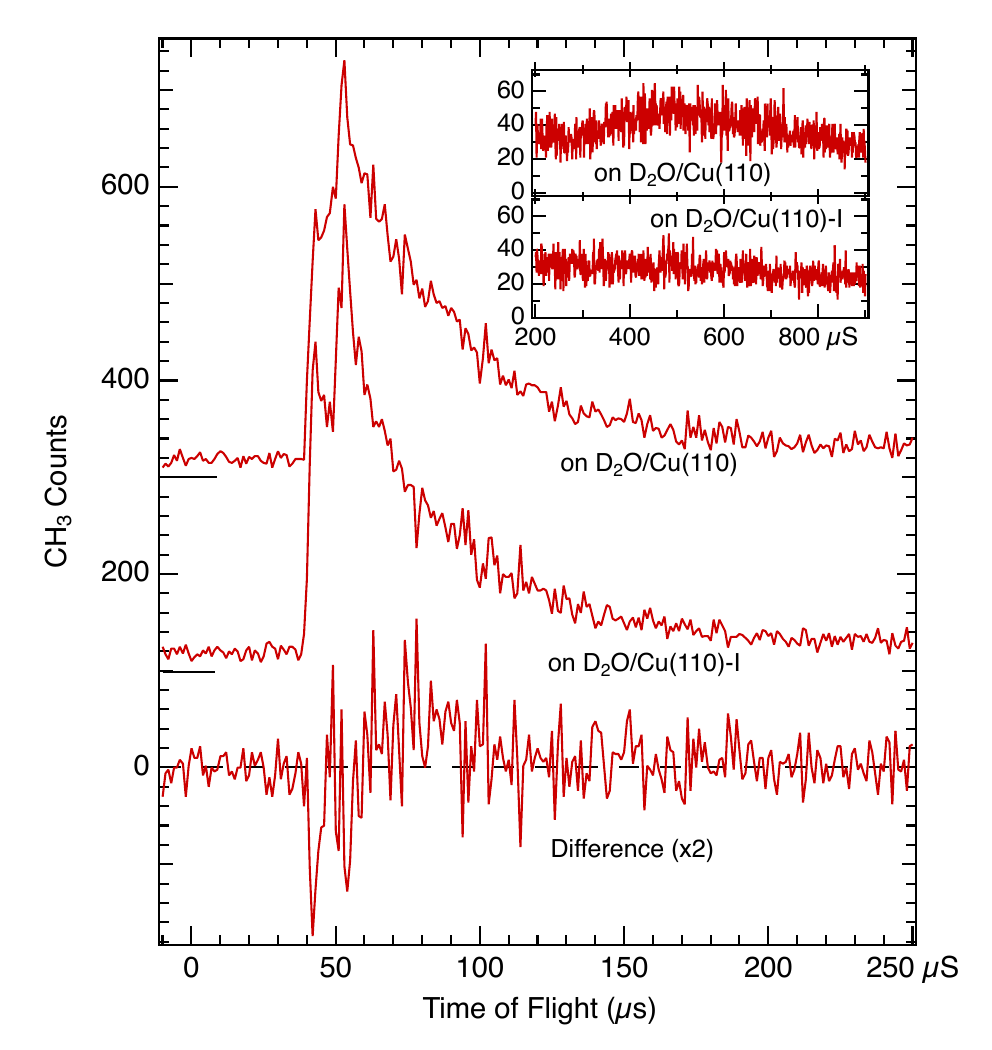}
\caption{\label{Fig_compare_Cu_CuI_monolayer} Comparison of TOF spectra for p-polarized light photodissociating 0.33ML CH$_3$I adsorbed on 1.0ML D$_2$O on Cu(110) (upper trace) and on Cu(110)-I (middle trace), and also showing the difference spectrum (bottom trace) between the two. The inset shows the region of the photodesorption feature, which is still observed on the Cu(110) substrate but not on the Cu(110)-I substrate. }
\end{figure}

Prior to the completion of the first D$_2$O layer, the yield of CH$_3$ photodissociation products is smaller and is consistent with earlier findings that CH$_3$I in contact with the Cu(110) substrate has its neutral photodissociation substantially quenched. The CH$_3$ photodissociation yield grows rapidly as the D$_2$O completes coverage of the copper surface, and then essentially constant CH$_3$ photodissociation yield is observed for thicker D$_2$O films (doses up to 120ML D$_2$O were studied). The dynamics of the photodissociation process are observed to shift as the amount of D$_2$O is varied-- Fig. {\ref{Fig_vary_D2O_tof}} shows a sequence of TOF spectra obtained from 0.33ML of CH$_3$I on 0ML to 15ML D$_2$O thin films. The spectra obtained for D$_2$O films in excess of 1.0ML thick are dominated by neutral photodissociation, and the TOF profiles are become essentially unchanging after 5ML D$_2$O coverage. For D$_2$O coverages below 1.0ML, we do observe contributions from both neutral photodissociation and also from a photodissociation pathway that is consistent with the DEA mechanism. More detailed TOF spectra for 0.33ML CH$_3$I with submonolayer D$_2$O coverages are shown in Fig. {\ref{Fig_low_dose_tof}}. The spectra of Fig. {\ref{Fig_low_dose_tof}} show a TOF feature centered near 80$\mu s$ flight time and which initially grows as the D$_2$O coverage is increased. Additional detail for these features is shown in Fig. {\ref{Fig_submonolayer_p_vs_s}}, in which TOF spectra are compared using p-polarized and s-polarized incident light. By switching to s-polarized laser light, dissociation via the $^3Q_0$ neutral pathway is suppressed in the spectrum as compared to the p-polarized spectrum, due to the vector nature of this absorption and the CH$_3$ photofragment detection geometry. Photon absorption by the substrate to create photoelectrons is relatively unaffected between the two polarizations, aside from differences in bulk light absorption coefficient ($A_p$/$A_s$$\approx$1.5 at 45$^\circ$ for Cu at 248nm). The data of Figs. {\ref{Fig_low_dose_tof}} and {\ref{Fig_submonolayer_p_vs_s}} support the assignment of the broad TOF feature centered around 80$\mu s$ to DEA of the adsorbed CH$_3$I. This feature  is not observed when a Cu(110)-I substrate is used. Also present in the same spectra of Fig. {\ref{Fig_low_dose_tof}} is a broad slow peak (between 250$\mu s$ and 850$\mu s$ flight times)-- this is ascribed to the photodesorption of intact CH$_3$I molecules from the surface via an electron stimulated mechanism\cite{Johnson:2000tr,Harris:1995dz}. These photodesorbed molecules are detected due to dissociation of the CH$_3$I in the QMS ionizer. The observation of the photodesorption feature can be used as a guide as to the absence or presence of significant quantities of photoelectrons that attach to the CH$_3$I molecules. In this way, we observe that the maximum photodesorption yield occurs for roughly 0.7ML D$_2$O (insensitive to the amount of CH$_3$I dosed subsequently), and diminishes when more D$_2$O is present. The DEA features reach a peak intensity for submonolayer amounts of preadsorbed D$_2$O, consistent with the peak photodesorption yield and is not easily discernable in the TOF spectra obtained using D$_2$O coverages greater than 1.0ML (e.g. in Fig. {\ref{Fig_tof_comparisons}}). To illustrate this, Fig. {\ref{Fig_compare_Cu_CuI_monolayer}} shows a comparison between TOF spectra obtained from 0.33ML CH$_3$I on 1.0ML D$_2$O dosed on top of a clean Cu(110) substrate and on an iodized Cu(110)-I substrate. The DEA feature seen at lower D$_2$O coverages on the Cu(110) substrate is not clearly evident in the top trace of Fig. {\ref{Fig_compare_Cu_CuI_monolayer}}. A difference spectrum (bottom trace in Fig. {\ref{Fig_compare_Cu_CuI_monolayer}}) between TOF spectrum from that on a Cu(110) substrate and from a Cu(110)-I substrate shows negative-going peaks at 43$\mu s$ and 53$\mu s$, indicative of somewhat more intensity in the CH$_3$I $^3Q_O$ dissociation pathways for the Cu(110)-I case, while a slight excess from 65$\mu s$ to 100$\mu s$ indicates a small contribution from DEA of CH$_3$I on 1.0ML D$_2$O on Cu(110). These findings demonstrate that DEA is at most a minor pathway for CH$_3$I photodissociation at 1.0ML and we find that this diminishes further for higher D$_2$O coverages. The photodesorption feature centered at 450$\mu s$ flight time is still observable in the spectrum of Fig. {\ref{Fig_compare_Cu_CuI_monolayer}} for the Cu(110) substrate but not on the Cu(110)-I substrate. These observations point to the smaller influence of photoelectrons on the CH$_3$I dissociation and photodesorption when the D$_2$O film thickness exceeds 1.0ML.

\begin{figure}
\includegraphics[scale=0.7]{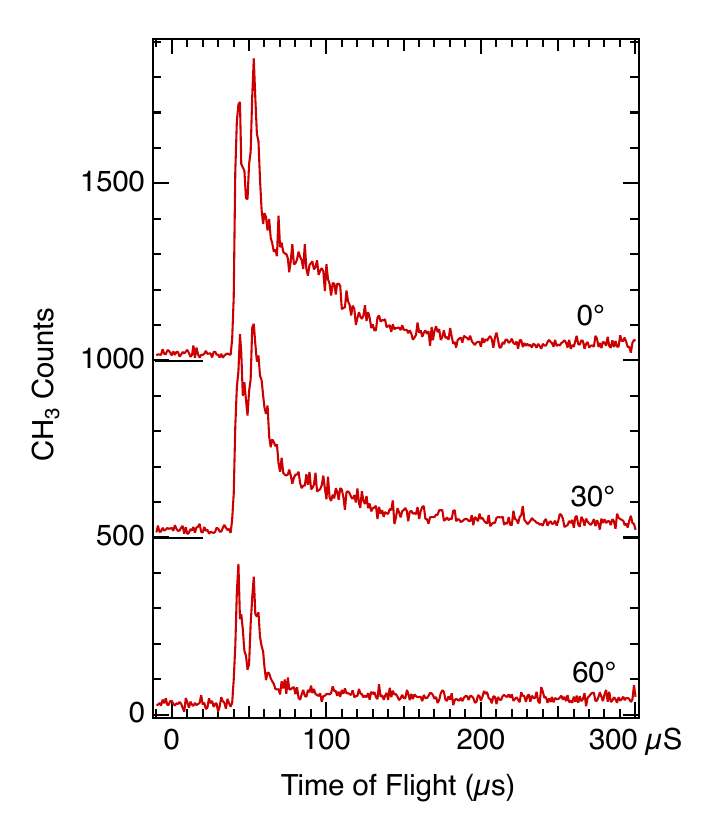}
\caption{\label{Fig_angular_tof_distns} Time of flight spectra for 0.33ML CH$_3$I adsorbed on 15ML D$_2$O/Cu(110) obtained at different angles from the surface normal, using p-polarized light. The proportion of the signal in the inelastic tail diminishes as the collection angle from normal increases. }
\end{figure}

Measurement of the angular distribution of CH$_3$ photodissociation products finds that the maximum intensity is in the surface normal direction, and decreases slowly with increasing angle away from the surface normal. For example, in the system of 0.33ML CH$_3$I adsorbed on 15ML D$_2$O/Cu(110), a fit to the angular yield of CH$_3$ photodissociation products using a fitting function of the form $\propto (\cos\theta)^N$ finds $N=1.9\pm 0.2$. However the TOF spectra {\it distributions\/} are observed to change significantly as a function of angle, as can be seen in Fig. {\ref{Fig_angular_tof_distns}}, which was obtained using p-polarized light for 0.33ML CH$_3$I on 15ML D$_2$O. As the angle from the surface normal is increased, the proportion of the CH$_3$ TOF spectrum in the inelastic ``tail'' of the spectrum diminishes. Comparison of the data obtained at $\theta=30^\circ$ with that at $\theta=60^\circ$ is particularly salient, as in both cases, the incident angle of the p-polarized laser light is $15^\circ$ from the surface normal (due to the fixed $45^\circ$ angle between the laser and TOF QMS axis). In these spectra it is clear that there is a lower proportion of the inelastically scattered CH$_3$ photofragments in the TOF spectrum obtained at $\theta=60^\circ$.

\begin{figure}
\includegraphics[scale=0.82]{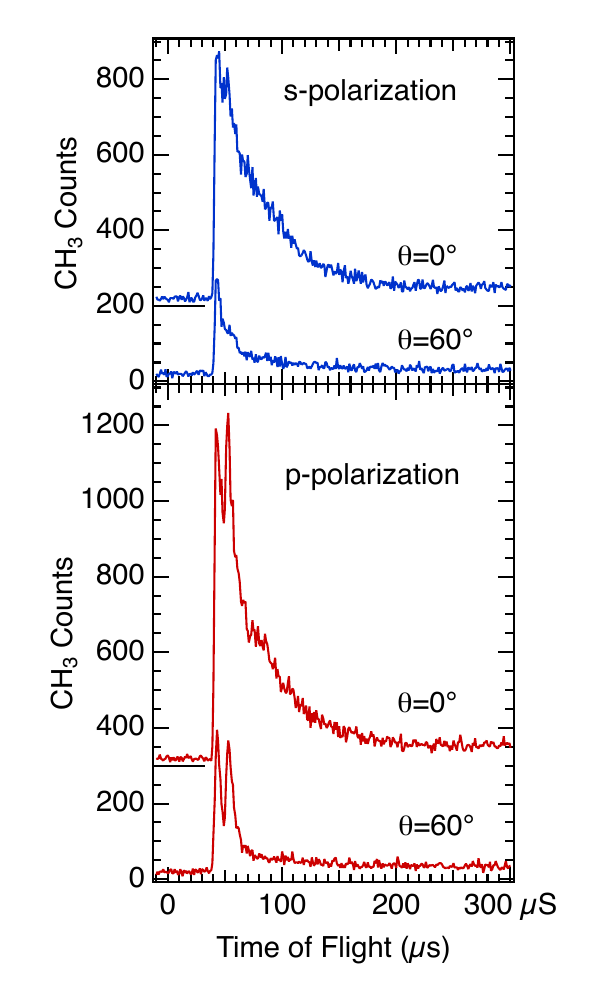}
\caption{\label{Fig_compare_angles_s_p} Time of flight spectra for 1.0ML CH$_3$I adsorbed on 15ML D$_2$O/Cu(110) obtained at $\theta=0^{\circ}$ and $\theta=60^{\circ}$ angles from the surface normal using s-polarized light (upper graphs) and p-polarized light (lower graphs).}
\end{figure}

Angular distributions obtained from 1.0ML CH$_3$I on 15ML D$_2$O/Cu(110) show the same basic features as for the submonolayer CH$_3$I case. Figure {\ref{Fig_compare_angles_s_p}} compares TOF spectra obtained at $\theta=0^{\circ}$ (surface normal) and $\theta=60^{\circ}$ obtained using p-polarized (lower spectra) and s-polarized (upper spectra) light. For both light polarizations, the magnitude of the inelastic tail in the TOF spectra diminishes as the angle is increased away from the surface normal. In the s-polarized data it can be seen that in addition to the fast onset peak at 43$\mu s$ (corresponding to dissociation via the CH$_3$ + I($^2P_{3/2}$) pathway), a small excess in signal is visible at 53$\mu s$. This observation supports the hypothesis that in addition to dissociation via the $^1Q_1$ excited state, s-polarized light also leads to dissociation via the $^3Q_0$ excitation, and that a substantial portion of CH$_3$ photofragments from this pathway are scattered toward the QMS detector.

By observing the diminution of the CH$_3$ photoyield as a function of the photon flux (assumed to reflect depletion of the adsorbed CH$_3$I), we find the depletion cross section to be $\sigma = 4\times 10^{-19}cm^2$ for 0.33ML CH$_3$I on 15ML D$_2$O/Cu(110) using p-polarized light. This cross section is comparable to that of the gas-phase absorption cross section at 248nm of about\cite{KellerRudek:2013wf} $8.5\times 10^{-19}cm^2$, and so is also consistent with the identified mechanism of neutral photodissociation\footnote{Some caution must be taken in interpretation of measured depletion cross sections in a heterogeneous chemical environment, as multiple factors can influence the measured yields and lead to non-exponential behaviour. In the present case, the depletion yields due to photon flux are well fit by exponential decreases, though this does not exclude the possibility of multiple operant mechanisms.}. The measured depletion cross sections are somewhat higher for monolayer and submonolayer coverages of D$_2$O, for example in the case of 0.33ML CH$_3$I adsorbed on 1.0ML D$_2$O/Cu(110) we measure the depletion cross section to be roughly twice that found for CH$_3$I on 5ML D$_2$O. The proportions of this increased depletion cross section that can be ascribed to neutral photodissociation, DEA and the observed photodesorption processes are unknown, as we do not know the relative ionization efficiencies for the CH$_3$ and CH$_3$I that lead to CH$_3^+$ in our QMS detector.

\section{Additional Discussion}

The TPD results for CH$_3$I adsorbed on D$_2$O multilayer films found evidence for two desorption features-- the $\alpha$ state at low coverage (exclusive up to $\sim$0.3ML) and the $\beta$ state at higher coverages. The data from photodissociation at varying CH$_3$I coverages (e.g. comparisons between Figs. {\ref{Fig_tof_comparisons}, \ref{Fig_angular_tof_distns}, \ref{Fig_compare_angles_s_p}}) found no appreciable differences in the TOF spectra, aside from increased CH$_3$ yield at higher CH$_3$I coverage. This would indicate that the structure of the adsorbed submonolayer CH$_3$I is essentially the same as that for the complete monolayer coverages. This observation is consistent with the structural model proposed in an earlier study of this system\cite{Perry:2007db}, as well as findings from another halogenated species, CCl$_4$, adsorbed on water\cite{Blanchard:1994ve,Sadtchenko:2000vi}. In this model, at low coverages the CH$_3$I is adsorbed on D$_2$O to form islands, and the associated $\alpha$ desorption feature in the TPD data is due to CH$_3$I monomers leaving the island edge before desorbing. At higher coverage, monomer desorption is no longer possible, resulting in the $\beta$ desorption feature from close-packed molecules that occurs at $\sim$10K higher temperature.

One surprising feature of the photodissociation findings is that the CH$_3$ photofragment yields from p- and s-polarized light data are similar in intensity (e.g. Fig. {\ref{Fig_p_s_pol_tof}}) even when only neutral photodissociation is the operant mechanism. As discussed above, this is apparently due to the substantial amount of inelastic scattering from CH$_3$I chromophores not oriented with the CH$_3$ moiety toward detector but which result in CH$_3$ exiting in the surface normal direction. It is difficult to extract the component TOF profiles from the spectra as the form of the inelastic profile is not known and resultant model fits are highly dependent upon the functions chosen. Nonetheless, it can be seen in the TOF spectra obtained at different angles (see Figs. \ref{Fig_angular_tof_distns} and \ref{Fig_compare_angles_s_p}) that the largest difference between these spectra is the magnitude of the contribution from the inelastic tail-- the magnitudes of the visible elastic peaks for the p-polarization appear to be similar at $\theta=0^{\circ}$ and $\theta=60^{\circ}$. It is an open question as to how the structure of the CH$_3$I adsorbed on D$_2$O ice leads to the observed profiles, but it seems likely that due to the CH$_3$I dipole moment and the dipolar nature of the ice I$_h$(0001) surface\cite{Henderson:2002wd}, that there is some degree of antiferroelectric ordering in the CH$_3$I adlayers. Hence one would expect a similar number of CH$_3$I oriented with CH$_3$ end ``up'' and ``down''. The photodissociation of downward pointing CH$_3$ could be a route to the generation of the observed substantial quantities of inelastic CH$_3$ photofragments. That a similar intensity of {\it elastic} CH$_3$ is observed in the surface normal direction as at $\theta=60^{\circ}$ also indicates that there is a wide range of initial polar angle orientations present in this system, although we cannot observe how these are distributed azimuthally (our QMS detector lies in the Cu(110) $[1\bar{1}0]$ azimuth).

\begin{figure}
\includegraphics[scale=0.49]{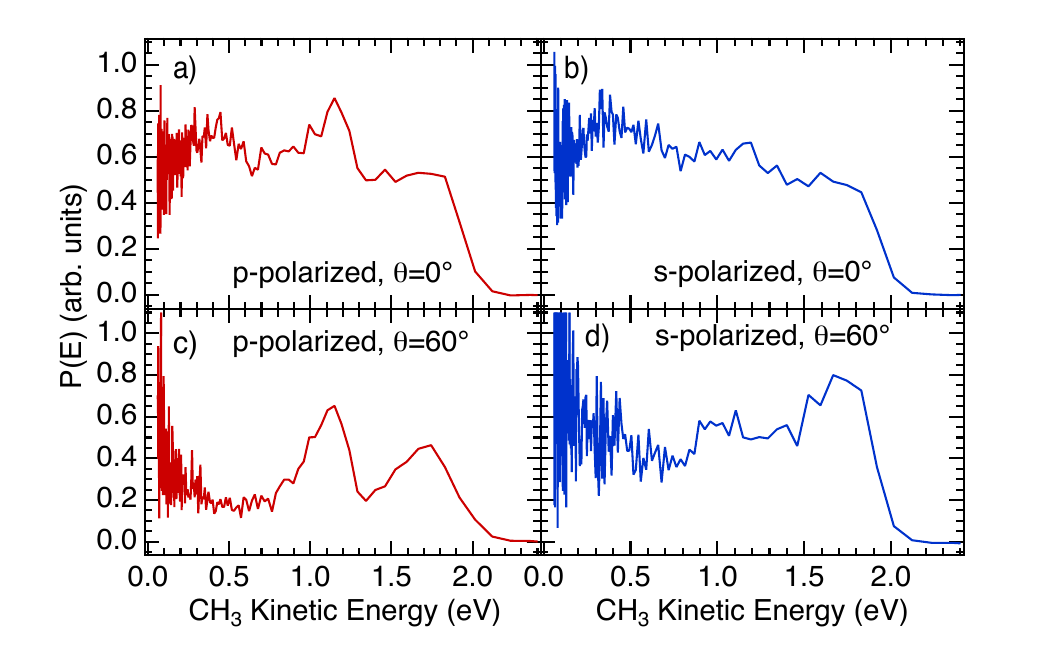}
\caption{\label{Fig_compare_angles_P_E} Time of flight data from Fig. \ref{Fig_compare_angles_s_p} (1.0 ML CH$_3$I adsorbed on 15 ML D$_2$O/Cu(110)) transformed to the energy domain to highlight the translational energy components present for the CH$_3$ photofragments. The vertical scale is proportional to observed flux of CH$_3$ at a given energy, but is not normalized in these plots. The peaks centered at 1.7 eV and 1.1 eV correspond to the fast CH$_3$ from the I and I$^*$ dissociation pathways. In the surface normal direction ($\theta$= 0$^\circ$), a peak centered at 0.4eV in (a) and (b) is diminished in spectra from $\theta$= 60$^\circ$ in plots (c) and (d). This feature is associated with CH$_3$ fragments that scatter multiple times before exiting the surface region.}
\end{figure}

A recent photodissociation study of CH$_3$I on thick H$_2$O layers\cite{DeSimone:2013hm} that detected the desorbing iodine (I, I$^*$, and I$_2$) found a substantial fraction of the iodine atoms leaving the surface with velocities that exceed the gas-phase limit. The largest proportion of the high kinetic energy iodine atoms were observed on low porosity amorphous ice, which is the same ice structure utilized in the thick D$_2$O layers of the present work. The most likely mechanism to produce these high speed iodine atoms are multiple collisions between the CH$_3$ photofragment and the I atom\cite{Tabares:1987tj}. This can occur following the photodissociation event if the CH$_3$ moiety is aimed downward toward the ice surface and then rebounds to recollide with the departing slow iodine atom. Although the study of DeSimone {\it{et al.}}\cite{DeSimone:2013hm} was performed at somewhat longer photodissociation wavelengths (260 nm and 290 nm) than the present work, photodissociation at these wavelengths follows the same A-band mechanism as outlined in the Introduction. An estimate of the energy scale involved would be to consider dissociation of CH$_3$I by 248 nm light to form CH$_3$ and I($^2P_{3/2}$). If the CH$_3$ fragment collides elastically with a rigid surface and then with the departing iodine atom, the resultant kinetic energy of the CH$_3$ would be 0.75 eV and that of the I would be 1.8 eV. Clearly this simple estimate would be the maximal possible energy and neglects excitations of both the surface and the CH$_3$ fragment as well as the I$^*$ dissociation pathway, which would have correspondingly lower kinetic energy release. The kinetic energy distributions for the departing CH$_3$ can be seen in the inset data of Fig. \ref{Fig_p_s_pol_tof} and also in Fig. \ref{Fig_compare_angles_P_E}, which shows the data from Fig. \ref{Fig_compare_angles_s_p} after being transformed to a plot of the flux versus kinetic energy. The peaks seen at 0.4 eV translational energy in Figs. \ref{Fig_compare_angles_P_E}(a) and (b) correspond to features centered around 90$\mu$s in the corresponding TOF data. That the observed CH$_3$ translational energy feature has a substantial fraction of the expected maximum from this simple collisional transfer model makes it reasonable to associate a significant contribution to the CH$_3$ TOF data from the same collisions that result in faster iodine photofragments that are observed in Ref. \onlinecite{DeSimone:2013hm}.

One motivation for the present work was a recent study of a similar system (CH$_3$I on D$_2$O/Cu(111)) which had identified a significantly enhanced photodissociation cross section for CH$_3$I diminution, and which was ascribed to DEA via photoelectrons solvated on the ice surface\cite{Sohn:2009tv}. The coverages of D$_2$O and CH$_3$I used in that study were reported as about 2.5ML for D$_2$O and somewhat more than 1ML for CH$_3$I. That study did not observe the dynamics of the dissociated CH$_3$I directly as the UV source used was a continuous emission Hg lamp, but inferred the mechanism by observation of changes to the $3d_{5/2}$ I XPS features and associated TPD results. The UV irradiance from an Hg lamp consists mainly of an intense band at 253.6nm (similar to the wavelength of the present study) and also a broader complex band at 290--330nm. In the present work, the CH$_3$ photofragment TOF distributions clearly indicate that at $\lambda$=248nm, the primary mechanism for photodissociation of surface CH$_3$I species on the D$_2$O thin films greater than 1.0ML thick is via neutral photodissociation. For D$_2$O submonolayer films, a DEA mechanism is identified but this does not dominate dissociation or lead to unusually large depletion cross sections for the coadsorbed CH$_3$I. It is possible that the DEA mechanism is of larger significance for CH$_3$I adsorbed {\it inside\/} the D$_2$O ice films and for which photodissociation does not produce fast CH$_3$ photofragments, but our TPD results (in agreement with \textcite{Perry:2007db}) find that only a small quantity of CH$_3$I appears to desorb coincident with  D$_2$O sublimation during TPD at 175K. Under the present experimental conditions, the CH$_3$I dosed on top of the D$_2$O thin films seems to be predominantly a surface species.

Our observations of the dynamics from photodissociation of CH$_3$I/D$_2$O/Cu(110) find that DEA is only a dominant pathway for D$_2$O coverages below roughly 1.0ML. Although it is difficult to untangle the contributions to photodissociation from the neutral processes and DEA, it is clear from the data of Figs. {\ref{Fig_low_dose_tof}} and {\ref{Fig_submonolayer_p_vs_s}} that the broad DEA translational energy distributions are more characteristic of a broad range of incident electron energies (characteristic of the broad energy range of hot photoelectrons created in the metal\cite{Weik:1993wa}) and not from a monoenergetic electron source\cite{Jensen:2008jb} that would be characteristic of electron transfer from an energetically well defined intermediate, such as solvated electrons on the D$_2$O surface. The CH$_3$ photodissociation results in the present work do not seem to support the suggestion that there is a large (i.e. orders of magnitude) level of enhanced photodissociation due to coadsorbed D$_2$O in this system. It is interesting to compare and contrast the findings in this work for CH$_3$I with that of CH$_3$Br on D$_2$O/Cu(110), in which significantly larger yields of CH$_3$ photodissociation by the DEA mechanism are apparent\cite{Jensen:2015dg}, though largely without the competing mechanism of direct photodissociation at $\lambda$=248nm. The much smaller DEA photodissociation yields observed for CH$_3$I as compared to CH$_3$Br adsorbed on D$_2$O/Cu(110) are surprising given that for these species adsorbed on dielectric thin films, the cross section for low energy DEA of CH$_3$I is more than twice as large\cite{Ayotte:1997th,Jensen:2008jv}. A possible explanation for this apparent discrepancy is the role of quenching. In these systems the adsorbed halomethane is in close proximity to the metal substrate and also potential electron traps in the D$_2$O\cite{Lu:2002da}, so it is plausible that the CH$_3$I$^-$ transient anion state has a larger overlap with, and thereby more rapid electron transfer (autoionization) with the D$_2$O and/or the Cu(110) substrate than for the CH$_3$Br$^-$ intermediate. Previous experience with CH$_3$I and CH$_3$Br DEA on heterogeneous molecular films\cite{Jensen:2008jb} suggests that substrate autoionization was much less effective when the temporary anion was formed on two or more layers of $n-$hexane, so in the present case of monolayer and submonolayer D$_2$O, the assignment of de-excitation mechanism is less clear.

That the yield from photodissociating CH$_3$I is found to increase as the submonolayer D$_2$O coverage increases is not surprising given that a variety of factors can lead to this observed behaviour. First, the neutral photodissociation of the CH$_3$I clearly begins to make a contribution to the yield even for submonolayer D$_2$O coverages-- the spectra of Figs. {\ref{Fig_low_dose_tof}} and {\ref{Fig_submonolayer_p_vs_s}} show the characteristic peaks (at 43$\mu s$ and 53$\mu s$ flight times) from the A-band photodissociation of CH$_3$I for D$_2$O coverages as low as 0.5ML. Secondly, the adsorption of water (D$_2$O or H$_2$O) decreases the surface workfunction-- by nearly 1.0eV upon completion of the monolayer for H$_2$O/Cu(110)\cite{Lackey:1989ve}. This decrease is somewhat larger than the magnitude of workfunction decrease due to the CH$_3$I/Cu(110) ($\Delta \Phi$=-0.53eV for 1.0ML CH$_3$I\cite{Johnson:2000tr}), but whatever the source, the decreased workfunction will play a role in increased flux of photoelectrons (both above and below the vacuum energy), which is known to modulate the DEA of adsorbed halomethane molecules\cite{Ukraintsev:1992ua,DixonWarren:1993di}. One further observation from Fig. {\ref{Fig_yield_vs_dose}} is that the photodissociation yield from CH$_3$I is very low for less than 0.3ML D$_2$O on Cu(110), and the slope of the yield curve is increased dramatically between 0.3ML and 1.0ML D$_2$O coverage. The change in slope in the 0.3ML region of Fig. {\ref{Fig_yield_vs_dose}} is likely due to the changes in D$_2$O adsorption structures at this coverage. For less than 0.3ML of water on Cu(110), the adsorption is in the form of 1-D chains in the [001] direction (perpendicular to the Cu(110) surface close-packed rows), but above this coverage, 2-D structures begin to grow\cite{Yamada:2006hb,Carrasco:2009eg}. The increased yield that occurs above 0.3ML D$_2$O coverage can be seen in the TOF spectra of Figs. {\ref{Fig_low_dose_tof}} and {\ref{Fig_submonolayer_p_vs_s}} to arise from both DEA as well as neutral photodissociation mechanisms. The increased photodissociation probability for CH$_3$I when the D$_2$O coverage is above 0.3ML is almost certainly due to the CH$_3$I adsorbing atop D$_2$O clusters and not in direct contact with the Cu(110) metal substrate, so that quenching is significantly reduced for the excited CH$_3$I for both the photon absorption and photoelectron capture dissociation mechanisms.



%
%

%

\begin{acknowledgments}
E.T.J. would like to thank the Natural Sciences and Engineering Research Council (NSERC) of Canada for financial support. E.R.M. and G.D.M. would like to thank NSERC for Undergraduate Student Research Awards which facilitated their tenures in the laboratory.
\end{acknowledgments}

\bibliography{Mechanism_for_CH3I_on_D2O}

\end{document}